\documentclass[prb,showpacs,twocolumn,preprintnumbers,amsmath,amssymb,floatfix]{revtex4}
\usepackage[dvips]{graphicx}
\usepackage[latin1]{inputenc}
\begin{document}
\draft

\newcommand{\lpc} {Li$_{0.5}$MnPc}
\newcommand{\lisi} {Li$_2$VOSiO$_4$}
\newcommand{\etal} {{\it et al.} }
\newcommand{\ie} {{\it i.e.} }
\newcommand{\aucr}{CeCu$_{5.9}$Au$_{0.1}$ }
\newcommand{\auaf}{CeCu$_{5.2}$Au$_{0.8}$ }
\newcommand{\aux}{CeCu$_{6-x}$Au$_{x}$ }
\newcommand{\ip}{${\cal A}^2$ }

\hyphenation{a-long}

\title{Coupling between \textit{4f} and itinerant electrons in SmFeAsO$_{1-x}$F$_x$ ($0.15\leq x\leq 0.2$) superconductors: an NMR study. }

\author{G. Prando$^{1,3}$, P. Carretta$^1$, A. Rigamonti$^1$, S. Sanna$^1$, A. Palenzona$^2$, M. Putti$^2$, M. Tropeano$^2$}
 \affiliation{$^1$ Department of Physics ``A.Volta", University of
Pavia-CNISM, I-27100, Pavia (Italy)}

\affiliation{$^2$ CNR-INFM-LAMIA and Università di Genova, I-16146 Genova (Italy)}

\affiliation{$^3$ Department of Physics "E. Amaldi", University of
Roma Tre-CNISM,  I-00146, Roma (Italy)}
\widetext

\begin{abstract}

$^{19}$F NMR measurements in SmFeAsO$_{1-x}$F$_x$, for $0.15\leq x\leq 0.2$, are presented. The nuclear
spin-lattice relaxation rate $1/T_1$ increases upon cooling with a trend analogous to the one already observed in
CeCu$_{5.2}$Au$_{0.8}$, a quasi two-dimensional heavy-fermion intermetallic compound with an antiferromagnetic
ground-state. In particular, the behaviour of the relaxation rate either in SmFeAsO$_{1-x}$F$_x$ or in
CeCu$_{5.2}$Au$_{0.8}$ can be described in the framework of the self-consistent renormalization theory for weakly
itinerant electron systems. Remarkably, no effect of the superconducting transition on $^{19}$F $1/T_1$ is
detected, a phenomenon which can hardly be explained within a single band model.

\end{abstract}

 \maketitle

\narrowtext

Although magnetism and superconductivity are often mutually exclusive phenomena they are observed to occur
simultaneously in several strongly correlated electron systems \cite{SCMag}. In the underdoped high-T$_c$
superconductors the presence of both phenomena suggested the onset of a microscopic phase separation within the
CuO$_2$ planes in magnetically ordered and superconducting regions \cite{Nied,Sanna1}. In those compounds also
rare-earth (RE) magnetism and superconductivity were found to coexist \cite{RE1}. A similar scenario was recently
found in Fe-based superconductors.\cite{FePS} At variance with hole-doped cuprates but similarly to electron-doped
ones,\cite{SmCuO,Marina2} in Fe-based superconductors RE \textit{f} electrons do not appear to be decoupled from
the Fermi sea. In fact, in superconductors of the so-called 1111 family, the reduction of the superconducting
transition temperature $T_c$ with pressure was explained in terms of a Kondo-coupling between \textit{f} and
conduction electrons \cite{Kondo}. Also the relatively large magnetic ordering temperatures of the RE ions
\cite{RE2}, in some cases exceeding 10 K, can hardly be explained without invoking a hybridization between
\textit{f} and conduction electrons, namely an RKKY coupling. Moreover,  the magnitude of the hyperfine
interaction between $^{75}$As nuclei and \textit{f} electrons in NdFeAsO$_{1-x}$F$_x$ \cite{Denis} suggests a
non-negligible coupling between \textit{f} and itinerant electrons. Even the magnitude of the Sommerfeld
coefficient in the specific heat indicates that the hybridization of the conduction electron wave functions with
RE \textit{f} orbitals leads to a renormalization of the effective electron mass.\cite{TropeanoA} Thus, it is
conceivable that the physics underlying the Fe-based superconductors of the 1111 family with a magnetic RE shares
some similarities with that of intermetallic heavy fermion compounds \cite{HF}.

In the following the study of the static and dynamic properties of SmFeAsO$_{1-x}$F$_x$ superconductors involving
\textit{f}-electrons will be discussed in the light of $^{19}$F NMR spectroscopy and nuclear spin-lattice
relaxation measurements. It will be shown that, remarkably, $^{19}$F nuclear spin-lattice relaxation rate $1/T_1$
is not affected by the superconducting transition. On the other hand, $1/T_1$ can be suitably described in the
framework of Moriya self-consistent renormalization (SCR) theory \cite{Moriya3} for weakly itinerant
two-dimensional (2D) antiferromagnets (AF). In fact, within this model one can explain both the temperature ($T$)
dependence of $^{19}$F $1/T_1$ in SmFeAsO$_{1-x}$F$_x$ and of $^{63}$Cu $1/T_1$ in CeCu$_{5.2}$Au$_{0.8}$, a 2D
heavy fermion AF \cite{CeCuAu}. The static uniform spin susceptibility derived from the  NMR shift follows a
Curie-Weiss law, as expected, with a relatively  large negative Curie-Weiss temperature.
\begin{figure}[h!]
\vspace{4.2cm} \includegraphics{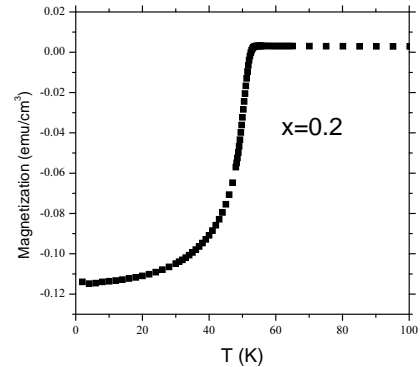}
   \caption{$T$-dependence of the field-cooled magnetization in SmFeAsO$_{0.8}$F$_{0.2}$,  for a magnetic field
   $H\simeq 5$ Oe.}
  \label{fig0}
\end{figure}

Polycrystalline Sm-1111 samples were synthesized in sealed crucibles of tantalum. \cite{Tropeano} This procedure
reduces F losses since it avoids the partial reaction of fluorine with the quartz vessel, so that it guarantees
that the doping content strictly scales with the nominal one, $x$, which is intended both as an upper limit to the
real content and as a sample label. The samples showed well defined superconducting transitions detected by means
of a superconducting quantum interference device (SQUID) magnetometer (Fig.\ref{fig0}). $\mu$SR measurements
performed in the $x=0.2$ sample show that the whole sample becomes superconducting below $T_c$ \cite{Sanna2}.

\begin{figure}[h!]
\vspace{6cm} \includegraphics{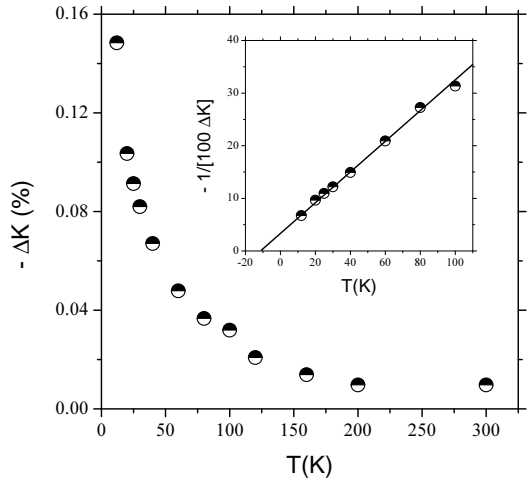}
   \caption{$T$-dependence of $^{19}$F NMR shift in SmFeAsO$_{0.85}$F$_{0.15}$,  for $H= 9$ Tesla.
   In the inset the inverse of the shift is reported as a function of $T$ in order to evidence the
   Curie-Weiss behaviour of the spin susceptibility.}
  \label{fig1}
\end{figure}

NMR measurements were performed by using standard radiofrequency (RF) pulse sequences. The intensity of the echo
signal was maximized by a $\pi/2-\tau-\pi/2$ solid echo pulse sequence and $^{19}$F NMR spectra were obtained from
the Fourier transform of the second half of the echo. The spectra were characterized by a negative shift, with
respect to $^{19}$F NMR signal in PTFE, (Fig.\ref{fig1}) which progressively increased upon cooling. The linewidth
was found to be weakly $T$-dependent above $T_c$, due to a small anisotropic dipolar hyperfine coupling. On the
other hand, a broadening was evidenced in the superconducting phase, due to the presence of the flux lines lattice
field distribution. The magnitude of this distribution is similar to the one detected by means of $^{19}$F NMR in
LaFeAsO$_{1-x}$F$_{x}$, \cite{Imai} however it appears to be significantly reduced with respect to the one
detected by $\mu$SR on the same SmFeAsO$_{0.8}$F$_{0.2}$ sample. The origin of this discrepancy will be discussed
elsewhere.

The $T$-dependence of the NMR shift $\Delta K$ is directly related
to the one of the static uniform spin susceptibility $\chi_s$. In
fact, one can write
\begin{equation}
\Delta K= \frac{A \chi_s}{g \mu_B N_A} + \delta
\end{equation}
where $A$ is $^{19}$F hyperfine coupling with the $f$-electrons, which dominates the response function, while
$\delta$ is the chemical shift. Hence, by plotting $\Delta K$ vs. $\chi_s$ estimated with a SQUID magnetometer,
leaving $T$ as an implicit parameter, one can estimate $A= -4.1\pm 0.3$ kOe. The $T$-dependence of $\Delta K$
indicates that $\chi_s$ follows a Curie-Weiss law with a Curie-Weiss temperature $\Theta= -11$ K. Remarkably
$\Theta$ is an order of magnitude larger than the one of SmBa$_{2}$Cu$_{3}$O$_{7}$, \cite{RE1} but close to the
one estimated for Sm$_2$CuO$_4$, where an indirect exchange coupling mechanism has been invoked.\cite{SmCuO} In
this latter electron-doped cuprate also the magnetic ordering temperature of Sm$^{3+}$ moments is very close to
the one found in SmFeAsO$_{1-x}$F$_{x}$. These observations indicate that the exchange coupling $J$ among
Sm$^{3+}$ magnetic moments in SmFeAsO$_{1-x}$F$_{x}$ cannot be justified in terms of a direct exchange mechanism
but rather suggests an indirect RKKY coupling.

\begin{figure}[h!]
\vspace{13cm} \includegraphics{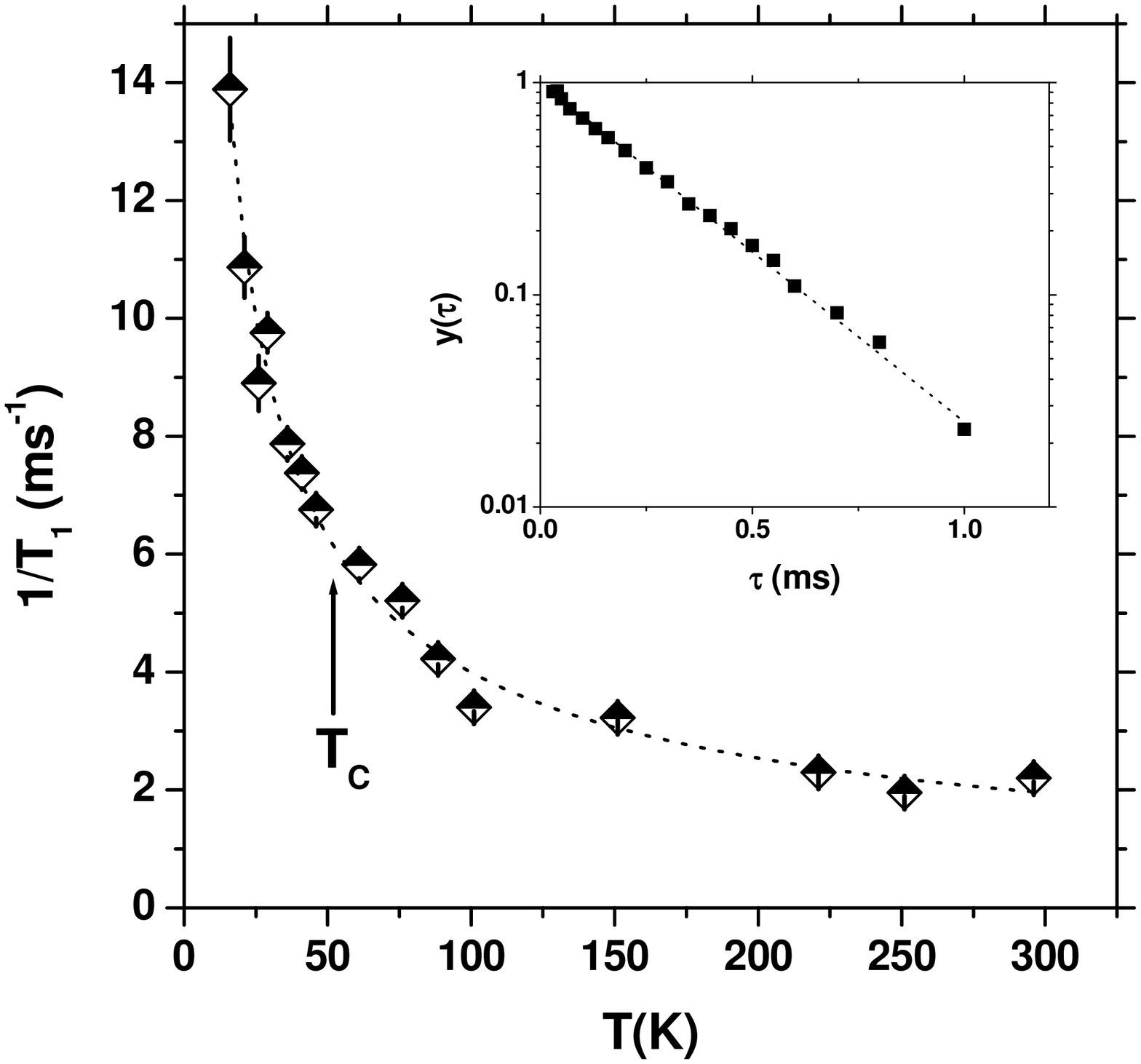} \includegraphics{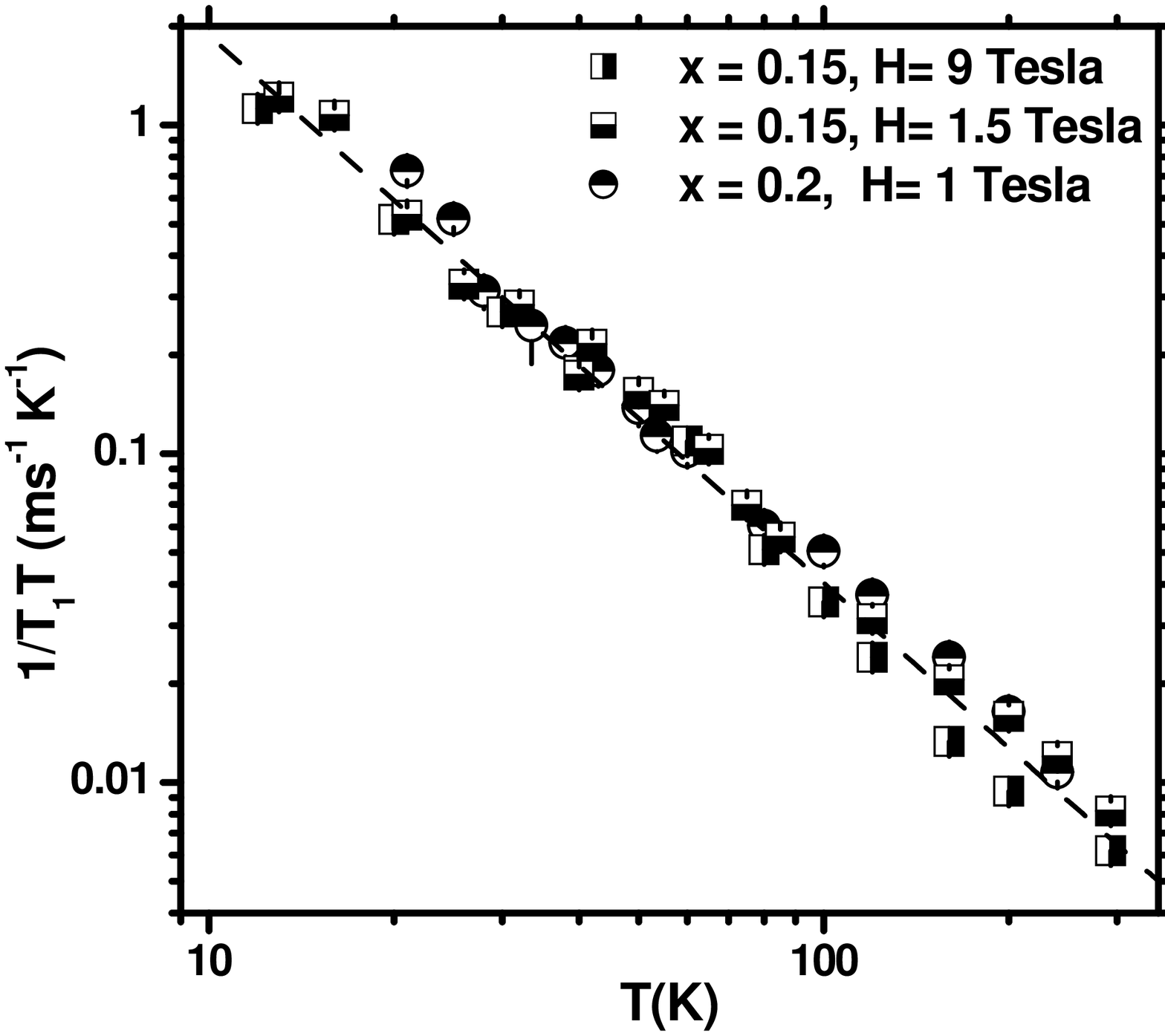}
   \caption{(Top) $T$-dependence of $1/T_1$ in SmFeAsO$_{0.8}$F$_{0.2}$, for $H= 1$ Tesla. In the inset
   a typical recovery law for the nuclear magnetization is reported. (Bottom) $T$-dependence of $1/T_1T$
   in the $x=0.15$ and $x=0.2$ samples, showing no significant field or $x$ dependence of $1/T_1$
   in this doping range. The dashed line represents the empirical power law $1/T_1T\sim T^{-1.6}$.}
  \label{fig2}
\end{figure}

Nuclear spin-lattice relaxation rate was derived from the recovery of the nuclear magnetization $m(\tau)$, after
$m(\tau)$ was set to zero by an appropriate excitation RF pulse sequence. The recovery of nuclear magnetization
$y(\tau)= 1-[m(\tau)/m(\infty)]$ was found to be a single exponential (Fig. \ref{fig2}), as expected for an
ensemble of $I=1/2$ nuclei with a common spin temperature and confirming the good sample homogeneity. The
$T$-dependence of $1/T_1$, derived by fitting the recovery laws with $y(\tau)= \exp(-\tau/T_1)$, is reported in
Fig. \ref{fig2}. One notices that $1/T_1$ increases with decreasing $T$ and, eventually, below about 10 K, the
short transverse relaxation time prevents the observation of $^{19}$F NMR signal. Remarkably no anomaly in the
$^{19}$F spin-lattice relaxation is detected at $T_c$ (see Fig. 3 at the top). The measurements, performed at
magnetic fields ranging from a few kGauss up to 9 Tesla, show that in the explored $T$ range $1/T_1$ is field
independent (Fig.\ref{fig2}).

Now we turn to the discussion of the $T$-dependence of $^{19}$F NMR $1/T_1$. First of all it is observed that
$^{19}$F spin-lattice relaxation rate in SmFeAsO$_{1-x}$F$_{x}$ is three orders of magnitude larger than in
LaFeAsO$_{1-x}$F$_{x}$ \cite{Imai}, which clearly indicates that $^{19}$F nuclei are probing low-energy
excitations involving Sm$^{3+}$ $f$ electrons. Such an enhancement was recently observed also for $^{75}$As
$1/T_1$ in 1111 superconductors with Pr or Nd.\cite{Kitaoka} Since $^{19}$F nuclei probe the correlated spin
dynamics within weakly coupled SmO layers one can at first try to justify the T-dependence of $1/T_1$ by
considering the $T$-dependence of the in-plane correlation length $\xi$ for a 2D AF with localized spins. For a
nuclear relaxation mechanism driven by spin fluctuations one can write
\begin{equation}
\frac{1}{T_1}= \frac{\gamma^2}{2} {k_BT} \frac{1}{N} \sum_{\vec q}
|A_{\vec q}|^2 \frac{\chi"(\vec q, \omega_R)}{\omega_R} \,\,\, ,
\end{equation}
with $\chi"(\vec q, \omega_R)$ the imaginary part of the dynamical spin susceptibility at the resonance frequency
$\omega_R$ and $|A_{\vec q}|^2$ the form factor describing the hyperfine coupling with the spin excitations at
wavevector $\vec q$. In the assumption that $|A_{\vec q}|^2$ does not filter out critical fluctuations, by using
2D scaling arguments one finds $1/T_1\propto \xi^z$,\cite{Sala} with $z=1$ the dynamical scaling exponent. For a
2D Heisenberg AF with localized spins one has $\xi\propto \exp(2\pi\rho_s/T)$, with $\rho_s\sim J$ the spin
stiffness \cite{CHN}. Since $J\sim \Theta\simeq -11$ K, it is difficult to justify within this model an increase
in $1/T_1$ starting at $T\simeq 200$ K $\gg |\Theta|$. The even more rapid increase of $\xi$ on cooling expected
for 2D Ising or XY systems would not explain the experimental results.

One could also consider that the excitations probed by $^{19}$F nuclei involve transitions among Sm$^{3+}$ crystal
field levels characterized by three doublets at energies of $E_1=0, E_2= 20$ and $E_3= 45$ meV \cite{Blundell}.
Then the relaxation processes would be Raman ones involving the exchange of energy $\hbar\omega_R$ between
Sm$^{3+}$ moments and the nuclei \cite{Prando}. Accordingly the $T$-dependence of $1/T_1$ is determined by the
Boltzmann factors describing the variation in the population of the crystal field levels.\cite{Prando2} Since in
the explored $T$-range $k_BT\ll E_3$ one can consider just the two low-energy doublets and one would find a
$T$-dependence characterized by an activated correlation time with an energy barrier $E_2$.\cite{Prando2} If one
tries to fit the data within this approach one would find a barrier one order of magnitude smaller than $E_2$,
showing that crystal field excitations cannot explain the spin dynamics.

On the other hand, as it was pointed out in the previous paragraphs, the presence of an indirect RKKY exchange
coupling would indicate a non-negligible hybridization between $f$ orbitals and the conduction band, a scenario
typically found in heavy fermion intermetallic compounds. Since no anomaly in $1/T_1$ is detected at $T_c$ these
conduction electrons  should not be or only weakly be involved in the pairing mechanism. This would be possible
only if different bands cross the Fermi surface, as it is the case here.\cite{Kondo,Band} Hence, the enhancement
of $T_c$ caused by Sm in the 1111 superconductors should be associated with a size effect only and not to a direct
involvement of $f$ electrons in the pairing mechanism.
\begin{figure}[b!]
\vspace{6.8cm} \includegraphics{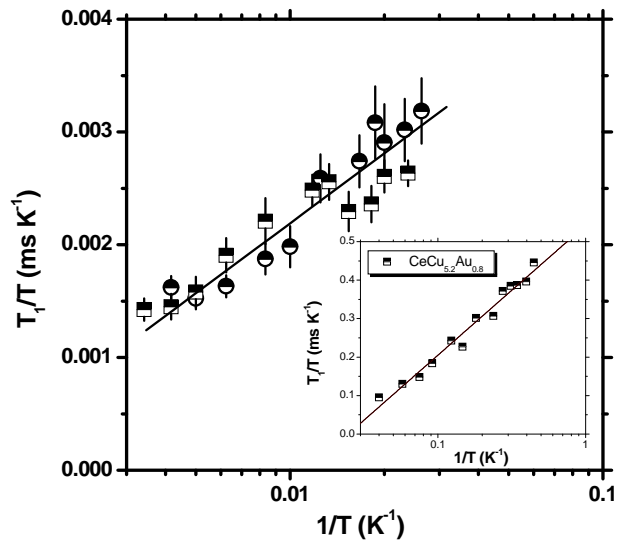}
   \caption{Semi-logarithmic plot of $^{19}$F $T_1/T$ vs. $T$ in  SmFeAsO$_{1-x}$F$_x$, for $x=0.15$
   (squares) and $x=0.2$ (circles). The solid line shows the best fit according to Eq. 7. In the inset the same type of plot
   is shown for $^{63}$Cu NQR $T_1/T$ vs. $T$ in  CeCu$_{5.2}$Au$_{0.8}$ (data from Ref.\onlinecite{NosCeCu}).}
  \label{fig3}
\end{figure}

It is interesting to notice that if one tries to fit the increase of $1/T_1T$ with a power law one finds
$1/T_1T\sim T^{-1.6\pm 0.1}$ (Fig. 3). This power law is nearly identical to the one found in CeFePO
\cite{CeFePO}, a compound with the same structure of SmFeAsO, where $^{31}$P $1/T_1T\sim T^{-1.5}$. In that
compound the behaviour of $1/T_1$ is consistent with the one of a weakly itinerant metal with a Fermi liquid
ground-state \cite{CeFePO}. Therefore, it is conceivable to analyze $1/T_1$ results for SmFeAsO$_{1-x}$F$_{x}$ in
the framework of the SCR theory developed by Moriya to describe weakly itinerant systems. Following Ishikagi and
Moriya \cite{Moriya1} one can write the dynamical spin susceptibility in terms of two characteristic parameters
$T_0$ and $T_A$ which characterize the width of the spin excitations spectrum in frequency and $\vec q$ ranges,
respectively. For antiferromagnetic correlations, as suggested by the negative Curie-Weiss temperature, one has
\cite{Moriya1}
\begin{equation}
\chi(q,\omega)= \frac{\pi T_0}{\alpha_Q T_A} \frac{1}{k_B2\pi T_0 (y + x^2) - i\omega\hbar} \,\,\,
\end{equation}
where $x=q/q_D$, with $q_D$ a Debye-like cutoff wave-vector, $\alpha_Q$ a dimensionless interaction constant  and
$y=1/2\alpha_Q k_B T_A \chi(0,0)$. Here $\chi(0,0)$ is the susceptibility per spin  in $4\mu_B^2$ units, with
dimensions of the inverse of energy, while $T_A$ and $T_0$ are in Kelvin. From the previous expression one can
derive $\chi "(\vec q,\omega_R)/\omega_R$ by taking the limit $\omega_R\rightarrow 0$, since $\hbar \omega_R$ is
well below the  characteristic energy of spin fluctuations. One can assume that the form factor $|A_{\vec
q}|^2\simeq A^2$ is almost $q$-independent, as expected for delocalized electrons. Then, by integrating $\chi
"(\vec q,\omega_R)/\omega_R$ over $\vec q$ in 2D, over a circle of radius $q_D$ centered at the AF wavevector
$Q_{AF}$, one derives
\begin{equation}
\frac{1}{T_1}= \frac{\gamma^2 A^2}{2} T \frac{\hbar}{4\pi k_B T_A T_0 \alpha_Q} \frac{1}{y(1+y)}
\end{equation}

Now, for correlated electron spins\cite{Moriya1} $y\ll 1$  and, by resorting to the expression for $y$ reported in
the paragraph above, one can simplify Eq. 4 in the form
\begin{equation}
\frac{1}{T_1}\simeq \frac{\gamma^2 A^2 }{4\pi} (\frac{T}{T_0}) \hbar \chi(Q_{AF})
\end{equation}
The $T$-dependence of $1/T_1$ in the previous equation is
determined by the one of $\chi(Q_{AF})$, which can be written in
terms of the in-plane correlation length $\xi$. Taking into
account the appropriate scaling and sum rules,\cite{Sala} one has
\begin{equation}
\chi(Q_{AF})= \frac{S(S+1) 4\pi \xi^2}{3k_BT \ln[4\pi\xi^2 + 1]}
\end{equation}
Since for $T\ll T_0$ the in-plane correlation length of this weakly itinerant metal should scale as
$\xi\sim\sqrt{T_0/T}$,\cite{CeCuCsi} by substituting this expression in Eq. 6 and then in Eq. 5 one has
\begin{equation}
\frac{1}{T_1}\simeq \frac{\gamma^2 A \Delta K}{2} \frac{\hbar}{\mu_B} \frac{1}{\ln[{4\pi T_0}/{T}]}\,\,\, .
\end{equation}
Finally, since for $T\gg \Theta$ the shift $\Delta K\propto 1/T$ one finds $({T_1}/{T})\sim \ln[{4\pi T_0}/{T}]$.
In order to check the validity of this expression we have first considered the $T$-dependence of $1/T_1$ in
CeCu$_{5.2}$Au$_{0.8}$, a heavy fermion intermetallic compound with 2D antiferromagnetic correlations which give
rise to a magnetic ground-state. In the inset of Fig. \ref{fig3} we report $^{63}$Cu $T_1/T$ for this compound
\cite{NosCeCu}. One notices that Eq. 7 nicely fits the data, with $T_0= 3.2 \pm 0.3$ K. In Fig. \ref{fig3} we
report the same plot for $^{19}$F nuclei in SmFeAsO$_{1-x}$F$_{x}$ for $x=0.2$ and $x=0.15$. In spite of the more
significant scattering in the data one notices that also in SmFeAsO$_{1-x}$F$_{x}$ the same logarithmic divergence
of $T_1/T$ is observed, with $T_0= 76\pm 15$ K.

In conclusion we have shown that in SmFeAsO$_{1-x}$F$_{x}$ the $T$-dependence of $^{19}$F $1/T_1$, driven by
\textit{f} electrons, can be explained by considering the low-energy excitations in SmO(F) layers as those of a 2D
weakly itinerant AF. This observation brings further support to a non-negligible coupling between \textit{f} and
conduction electrons in the superconductors of the 1111 family and to an active role of $f$ electrons in
determining the electronic properties. The absence of any anomaly in $1/T_1$ at $T_c$ suggests the presence of
different bands crossing the Fermi surface, not all of them significantly involved in the pairing mechanism.



\end{document}